\documentclass[11pt,titlepage]{article}

\usepackage{setspace}
\usepackage{amsmath,lastpage,bm,textcomp}
\usepackage{graphicx}
\usepackage[round,numbers,sort&compress]{natbib}
\usepackage{float}
\usepackage{authblk}
\usepackage{color}

\title{Domain-domain interactions in Filamin A (16-23) impose a hierarchy of unfolding forces}

\author[*]{Tianyou Xu}
\author[*]{Herbert Lannon}
\author[*]{S\'ebastein Wolf}
\author[+]{Fumihiko Nakamura}
\author[*]{Jasna Brujic\thanks{Corresponding author. 608 Meyer Hall, 4 Washington Place, New York, NY, 10003, USA, Tel:~(212)998-3586, Email: jb2929@nyu.edu}}
\affil[*]{Department of Physics and Center for Soft Matter Research, New York University, 4 Washington Place, New York, NY, 10003, USA}
\affil[+]{Translational Medicine Division, Department of Medicine, Brigham and Women's Hospital, Harvard Medical School, 75 Francis Street, Boston, MA, 02215, USA}

\date{}

\begin{document}
\maketitle
\abstract{The quaternary structure of Filamin A (FLNa) 16-23 was recently shown to exhibit multiple domain-domain interactions that lead to a propeller-like construction. Here we present single molecule force spectroscopy experiments to show a wide variety of mechanical responses of this molecule and compare it with its linear counterpart FLNa 1-8. The compact structure of FLNa 16-23 leads to a broad distribution of rupture forces and end-to-end lengths in the force-extension mode and multiple unraveling timescales in the force-clamp mode. Moreover, a subset of force-extension trajectories reveals a mechanical hierarchy in which the rupture of domain-domain interactions at high forces ($>$200~pN) liberates the unfolding of individual domains at low forces ($\sim$100~pN). This mechanism may also explain the order of magnitude difference in the rates of the biexponential fits to the distribution of unfolding dwell times under force-clamp. Overall, FLNa 16-23 under a force of 100 pN is more compliant than the linear FLNa 1-8.  Since a physiological role of FLNa is to crosslink actin filaments, this range of responses allows it to accommodate a broad spectrum of forces exerted by the cell and its environment.

}

\clearpage

\section*{Introduction}

Many proteins assemble into supramolecular structures in order to perform their biological function. For example, pilin monomers assemble into pili fibers that regulate bacterial locomotion \citep{pili1, pili2, pili3}, ankyrin repeats assemble into superhelical spirals that may influence neuronal function of the inner ear \citep{Lee}.  ~Conversely, some proteins aggregate into dysfunctional amyloid fibrils that impact the development of neurodegenerative diseases~\citep{Chiti2006}. These complex structures have been characterized by single molecule force spectroscopy unveiling a diversity of mechanical behaviors~\cite{amyloid_force, fibrils}. In the case of ankyrin repeats, $\alpha$-helical domain-domain interactions lead to structures that are much more mechanically stable than the individual folds, inducing a hierarchy in their force response \citep{Lee}. Similarly, dimerization through $\beta$-sheet Ig domains Z1Z2 in titin leads to rupture forces as high as 700~pN that are significantly stronger than their unfolding force of $\sim$170pN~\citep{julio_titin}. By contrast, stretching linear polyprotein chains, such as the tandem repeats in titin and ubiquitin, results in the unfolding of the individual domains in the order imposed by their mechanical stability~\citep{titin, ubiquitin}. Note that this order may be disrupted by the presence of stable intermediates within individual domains, as in the case of fibronectin~\citep{fibro}. More generally, the specific force response of a given system reveals whether structural hierarchies are present and the nature of the underlying interactions. 

In this study, we investigate the polyprotein FLNa, which exhibits a distinct mechanical architecture~\citep{tweezers}. It is an actin-binding protein whose function is to crosslink the intracellular network of F-actin and regulate the remodeling of the cytoskeletal infrastructure~\citep{Stossel2001, Nakamura2011}. To achieve this function, it has been proposed that FLNa plays a role in mechanotransduction: it serves as an integrator capable of detecting mechanical changes and triggering appropriate biochemical reactions~\citep{Nakamura2011,Addario,mechanoprotection,actin_mechanics1}. In fact, FLNa has been reported to interact with over 30 cellular proteins, which implies that its functionality is versatile and contributes to many signaling pathways~\cite{Feng2004}. FLNa is therefore implicated in a wide range of physiological functions that are affected by the mechanical resilience of the protein. Our aim is to unravel the diverse responses of this molecule to a stretching force in order to reveal structural features within its architecture.

FLNa is a long, rod-like polyprotein consisting of 24 repeat immunoglobulins. This FLNa monomer forms a V-shaped dimer complex that bridges two actin filaments as shown in Fig.~\ref{fig:one}. The Ig domains 1-15 linearly bind to one actin filament and the clustered domains 16-24 bridge the gap between two adjacent actin filaments~\citep{nakamura}. These domains interact in a pairwise manner and assemble into propeller-like structures~\citep{domain_1, domain_2}. The inset shown in Fig.~\ref{fig:one} illustrates the crystalline loop structure of FLNa 19-21, which specifically involves hydrogen bonding between the beta pleated A-strand in FLNa 20 and C-D strands in FLNa 21~\citep{pdb}. The unbinding of domains 20 and 21, which also occurs between domains 16-17 and 18-19, is predicted by molecular dynamics simulations to require $\sim$50~pN of force~\citep{binding,Heikkinen2009,Pentikainen2009}. Interestingly, recent force-ramp experiments using magnetic tweezers reveal that FLNa 16-23 unravels at significantly lower pulling forces than sub-segments of linear FLNa 1-15 containing equivalent numbers of FLNa Ig repeats~\citep{tweezers}. The molecular mechanisms by which the unraveling occurs and the effect of domain-domain interactions on the mechanical response of the protein remain open questions. 

Using the atomic force microscope (AFM) here we investigate the force response of FLNa 1-8 and 16-23 over a broad range of forces in the force-extension mode, as well as the kinetics of unraveling under a constant stretching force. We collect a large statistical pool of data to show that both linear chain behavior and higher-order complexities coexist in the polyprotein. Features in the experimental distributions of rupture forces, end-to-end  lengths and dwell times are then used to elucidate the mechanical hierarchies and their possible biological significance.  

\section*{Materials and Methods}
Fusion proteins were expressed in Sf9 cells (5108 cells) using Bac-to-Bac system (Invitrogen) in accordance with manufactures' instructions and the expressing cells were harvested 72 h post-infection as previously described \citep{nakamura}. After washing with PBS, the cells were lysed in 40 mL of 20 mM sodium phosphate, pH 8.0, 1\% Triton X-100, 300 mM NaCl, 20 mM imidazole, 1 mM mercaptoethanol, 2 mM PMSF, 10 g/ml aprotinin, and 10 g/ml leupeptin at $4^{\circ}$C. The extracts were centrifuged at 20,000g for 30 min at $4^{\circ}$C and loaded onto a Ni-NTA column (3 ml, Qiagen). The column was washed with washing solution (20 mM sodium phosphate, pH 8.0, 20 mM imidazole, 1 mM mercaptoethanol, 300 mM NaCl, 0.1\% Triton X-100) and bound recombinant proteins were eluted with 20 mM sodium phosphate, pH 8.0, 300 mM imidazole, 1 mM mercaptoethanol. Purified proteins were concentrated using an Amicon Ultra-15 (Millipore) with a molecular weight retention of $>$5,000 Da and gel-filtered on Superdex 200 (10/300, GE Healthcare) column equilibrated with 50 mM Bicine-NaOH, pH 8.3 and 0.1 mM mercaptoethanol. There were no signs of aggregation during these processes. Finally, we dialyze the protein solution against PBS buffer solution at pH=7.4. Recombinant proteins were stored at $-80^{\circ}$C.

AFM measurements were conducted using a custom-built apparatus consisting of a modified Digital Instruments detector head (AFM-689, Veeco Instruments, Santa Barbara, CA) and a three dimensional piezoelectric translator, with a range of 5 $\mu$m and a resonant frequency of $\sim$ 10 kHz (P-363.3CD, Physik Instrumente, Karlsruhe, Germany). Laser position and alignment were performed using a photo-diode at 100 kHz (Schafter Krichhoff). Details of the AFM and its operation modes have been described elsewhere~\citep{Nature1998, Science2004}. Silicon nitride cantilevers (MLCT, Bruker Probes, Camarillo, CA) were used, with spring constants calibrated using the equipartition theorem to be $\sim20-25$pN/nm. An aliquot of 10-15 $\mu$L of poly-protein (0.10 mg/mL, buffer pH 7.4) was deposited on an evaporated gold surface and immobilized for 10 minutes. All experiments were performed at room temperature. Poly-protein chains were picked from the surface by non-specific binding the the AFM tip. Chains were then pulled at a constant velocity or constant force, depending on the mode of operation. The experiment is regulated by a PID controller and run using software written in Igor PRO. 

In our experiments, a single molecule is characterized by: a sequence of at least three peaks in the force extension mode or three steps in the force-clamp mode, a total length of the trajectory that does not exceed the fully extended FLNa molecule $L_{tot}\sim240$nm and a time course in the force-clamp trajectories that is at least 1.5 seconds long. We also exclude trajectories with considerable drift (larger than 20pN) or trajectories that do not exhibit sharp peaks or staircases. It is therefore highly unlikely that the reproducible features in the data are attributable to spurious interactions.   

The quantitative analysis of the traces involves fitting the worm-like chain (WLC) model defined by,
\begin{equation}
  F(x)=\frac{k_{B}T}{L_p}(\frac{1}{4}(1-\frac{x}{L_c})^{-2}-\frac{1}{4}+\frac{x}{L_c})
\end{equation}
where $x$ is the extension in the end-to-end length of the molecule, $L_p$ is the persistence length and $L_c$ is the contour length \citep{wlc,wlc2}. For the fitting, $L_p$ is fixed to be the size of one amino acid of $\sim0.37$~nm. In the case of force-extension data, we measure the distance $\Delta L_{FX}$ from the difference in $L_c$ estimated for two consecutive peaks in the sawtooth pattern. This $\Delta L_{FX}$ corresponds to the contour length of the molecule $L_c$ minus its folded length $L_f$. Since $L_c$ is given by the number of amino acids times the size of an amino acid and $L_f$ of Ig domains is estimated to be $\sim3.7$~nm~\citep{pdb}, we can deduce whether rupture events correspond to the unfolding of individual domains. In the case of force-clamp data, the step-sizes correspond to the end-to-end length change from an extended, folded state to an unfolded state under the constant, finite force. The step size $\Delta L_{FC}$ can be estimated using the WLC model at the applied force of 100~pN minus the folded length $L_f$ of the Ig domain.

\section*{Results and Discussion}

\begin{figure}
\begin{center}
\includegraphics[width=1.0\textwidth]{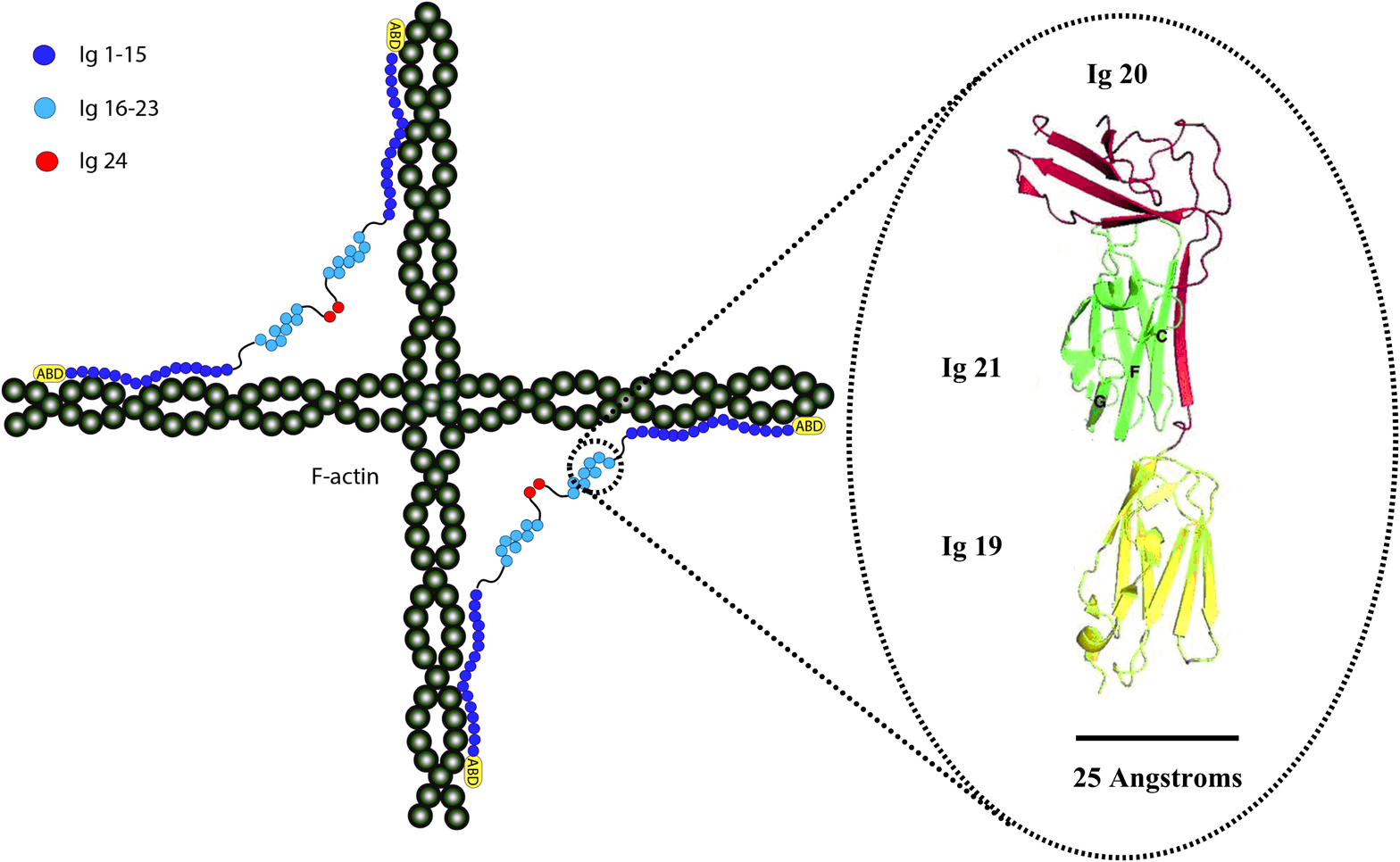}
\caption{}
\label{fig:one}
\end{center}
\end{figure}

\begin{figure}
\begin{center}
\includegraphics[width=1.0\textwidth]{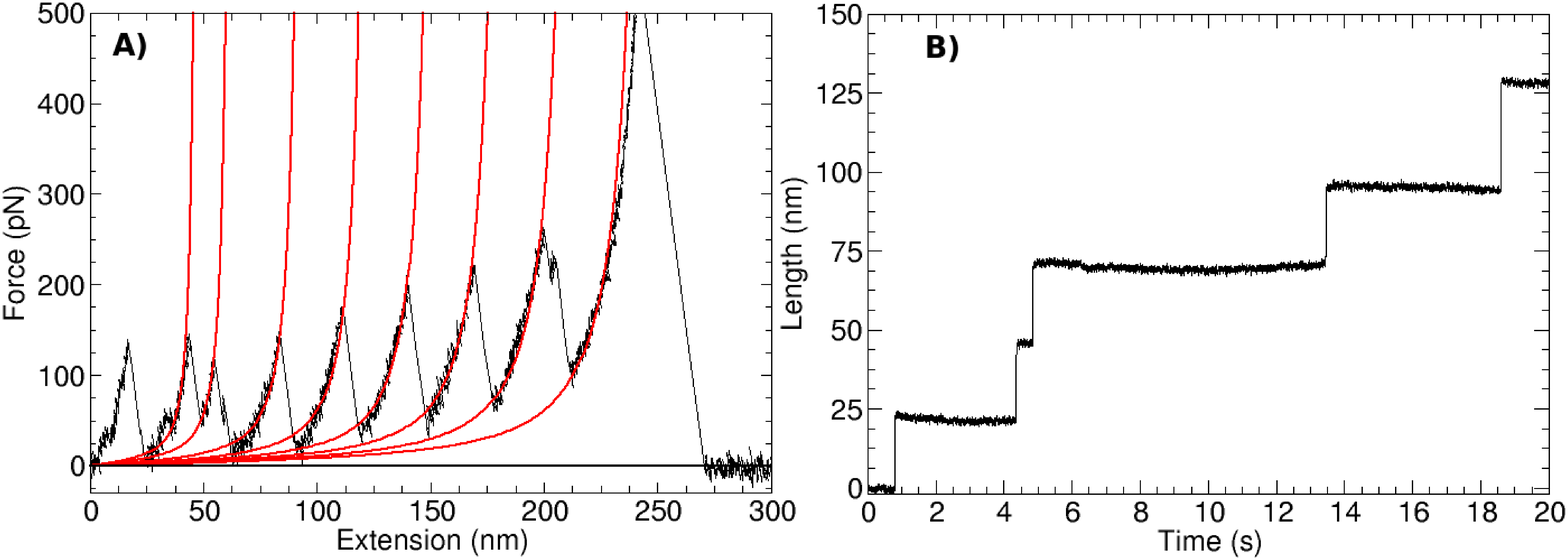}
\caption{}
\label{fig:two}
\end{center}
\end{figure}

In Fig.~\ref{fig:two},~A and B we show trajectories of FLNa 1-8 under a constant velocity of 400~nm/s and a constant force of 100~pN. Domains 1-8 behave as linear polyprotein chains, as depicted in Fig.~\ref{fig:one} and observed in previous AFM studies of FLNa~\citep{Furuike}. Each peak of the sawtooth represents the unfolding of a domain in the chain, while the last peak is the detachment of the chain from either the cantilever tip or the gold surface. The WLC fits to the sawteeth reveal distances in the range $\Delta L_{FX}=27-33$~nm, similar to the predicted values shown in Table~\ref{tab:table} \citep{uniprot}. Moreover, the steps in the unfolding staircase in the range $\Delta L_{FC}=25-27$~nm correspond to the predicted extensions of Ig domains listed in the table. These results are in agreement with the electron micrographs in~\cite{nakamura}, which show a contour length consistent with a linear construct. Furthermore, previous force studies using magnetic tweezers~\cite{tweezers} and AFM~\cite{Furuike} also observed reproducible sawtooth patterns in agreement with the unfolding of individual domains. The average unfolding force of~$\sim200$~pN coincides with that of the 27th Ig domain (I27) in titin~\citep{i27}. Using the unfolding rate $a_0=3.3x10^{-4}s^{-1}$ in the absence of force and the distance to the transition state $dx=0.25nm$ of I27~\citep{i27} in the Bell model~\citep{Bell}, $a=a_0exp(Fdx/kT)$, predicts an unfolding timescale $a=0.15~s^{-1}$ at the constant force of $F=100$pN and $kT=4.1$pNnm. Indeed, a rough estimate of the timescale of the trajectory in Fig.~\ref{fig:two} B is on the same order of magnitude as the predicted one.

By contrast, in Fig.~\ref{fig:three} we show a gallery of responses to force in FLNa 16-23, which is expanded upon in the Supplementary Materials Figs. S1-3. Typical trajectories in FLNa 16-23 can be classified into two categories: those that exhibit a broad distribution of end-to-end lengths at similar rupture forces, shown in Fig.~\ref{fig:three},~A and B, and those that display a reverse mechanical hierarchy in which the strong interactions break before weak ones, shown in Fig.~\ref{fig:three},~C and D. More specifically, Fig.~\ref{fig:three} A shows rupture events at $\sim150$~pN where seven out of the nine peaks have $\Delta L_{FX}$ outside of the range of the unfolding of individual domains. This result is in agreement with the broad range of step sizes in the analogous constant force trajectory at $100$~pN in Fig.~\ref{fig:three} B. The lower unfolding force of 150pN is similar to that of native Fibronectin~\cite{fibro}, which has a similar structure to Ig domains in that it is a 7-strand beta-sandwich. Analysing the data in~\cite{fibro} using the theoretical formalism in~\cite{Hummer2003} predicts $a_0=1.1\times10^{-4}s^{-1}$, $dx=0.38$nm, and an unfolding rate at $100$pN constant force in force-clamp of $a=1.2s^{-1}$. This rate constant is similar to the time course of the trajectory in Fig.~\ref{fig:three}B. This unfolding rate is an order of magnitude faster than the unfolding staircase in Fig.~\ref{fig:two}B. 

In the cases shown in Fig.~\ref{fig:three},~A and B the FLNa 16-23 is significantly weaker than its linear counterpart FLNa 1-8. Since the contour length of the whole FLNa 16-23 is $\sim240$~nm, which is only slightly longer than the observed trajectory in Fig.~\ref{fig:three} A, these events must involve the unfolding of the individual Ig domains. A possible explanation for the weaker unfolding as well as the breadth of the released lengths is that the direction along which the force is applied to each domain is imposed by the packing of the Ig domains within the polyprotein chain, as shown in the propeller-like structure proposed in~\cite{domain_1,domain_2} as well as in Fig.~\ref{fig:one}. Indeed, changing the linkage between a tandem of linear ubiquitin monomers was found to change the rupture forces and released lengths according to the pulling direction~\citep{Carrion-Vazquez2003}. 

The second category of typical FLNa 16-23 responses under the same experimental conditions reveals strong rupture events followed by weak ones in the force-extension mode and multiple unraveling timescales in the force-clamp mode, as shown in Fig.~\ref{fig:three},~C and D. In particular, Fig.~\ref{fig:three} C shows three rupture forces at $\sim200$~pN that are approximately twice as strong as those that follow.  The inset shows that the frequency of trajectories with $n$ liberated low force peaks decreases with $n$. Similarly, Fig.~\ref{fig:three} D shows the constant force unraveling of three strong events over 3 s, followed by  the rapid succession of three weak events in only 0.7 s. These observations suggest that the weak events are initially hidden from the applied force within the protein superstructures and only subsequently released. This interpretation is plausible given the hierarchical packing structure of FLNa domains 16-23 proposed in Fig.~\ref{fig:one}.   

It should be noted that the fitting with the WLC model is not as successful in the trajectories of FLNa 16-23 as in the linear sawtooth patterns of FLNa 1-8. In the Supplementary Materials Figs. S5-7 we show that the fits can be improved by the addition of a Hookean spring. However, this result does not change the measured values of $\Delta L_{FX}$ beyond the error estimates in the fitting, as shown in the Supplementary Table S1. In addition, allowing the persistence length to vary in the fits increases the error in fitting $L_c$, but does not change the mean value, as shown in the Supplementary Table S2.

Next, we analyze the statistics of all the collected trajectories. The histograms of rupture forces and distances in $\Delta L_{FX}$ for FLNa 16-23 at constant velocity reveal notable differences compared to the results found in previous studies on FLNa \citep{Furuike}. The force distribution ($N=850$ events) in Fig.~\ref{fig:four}~A spans a much broader range than that previously found in FLNa Ig domains (shaded histogram) and reaches values higher than the unfolding force of any known mechanically stable protein ~\citep{stable}. These rupture forces on the order of 700~pN are comparable to the unbinding of the quaternary structure of Ig domains Z1Z2 in titin~\citep{julio_titin}. In addition, the low force regime probed by magnetic tweezers broadens the rupture force distribution all the way down to 5~pN ~\citep{tweezers}. The inset shows that a power law tail captures the experimental distribution. This scale-free force response to the extension of the end-to-end length makes this protein uniquely adaptable to a wide range of external perturbations. Likewise, the distribution of $\Delta L_{FX}$ in Fig.~\ref{fig:four} B is much broader than the shaded histogram found in Ig domains, indicative of the proteins' complex response to force. Note that the lengths span from extensions of short intermediate or domain-domain ruptures of $\sim5$~nm up to the extension of multiple domains of $\sim90$~nm. 

Despite the broad distributions observed in FLNa 16-23, the probability map of $\Delta L_{FX}$ as a function of the rupture force in the inset reveals a peak in the region that corresponds to the linear unfolding lengths of Ig domains ($\sim28$~nm) at rupture forces between $100-150$pN, in agreement with the unfolding forces of weaker domains in FLNa. A filtered histogram of $\Delta L_{FX}$ in Fig.~\ref{fig:four} C for only those events that occur after a rupture force that is at least twice as high as their rupture force isolates the same unfolding peak at $28$nm. The fact that the probability map for this reduced data set is consistent with domain unfolding, as shown in the inset,  means that the rupture of higher order structures linearizes the polyprotein chain and liberates Ig domain unfolding.  
 
Force-clamp data at $100$~pN on the two protein constructs reveals a distribution of step sizes for FLNa 16-23 ($N=830$ events) in Fig.~\ref{fig:four} D and a single peak at $\sim22$~nm for FLNa 1-8 ($N=372$ events) in the inset. These results are consistent with force extension distributions for the two constructs. The broad distribution of step sizes in FLNa 16-23 exhibits a peak at $\sim20$~nm, which is roughly consistent with the unfolding lengths of the individual domains. While the force-extension mode probes structures with broad interaction energies, pulling at a constant force isolates only those that are susceptible to the applied force in the time window of the experiment, i.e. $5$ seconds. For this reason, it is harder to distinguish unfolding events from domain-domain ruptures according to the order in which they occur. Instead, we consider the timescales on which the unraveling of the end-to-end length occurs.

To study the kinetics of the unraveling processes of FLNa 1-8 and 16-23 we analyze the distribution of dwell times for each step in the staircases observed under force-clamp. The cumulative distributions obtained from different responses of the same molecule are useful in that they give the macroscopic, viscoelastic response of the molecule. When the actin cytoskeleton is placed under tension in vivo, each of the FLNa molecules that is crosslinking the actin will be exposed to force along different directions. Averaging over all the individual responses in our experiments allows us to probe the global response of each construct to a stretching force.

In the case of linear FLNa 1-8, the fact that the step sizes are uniform implies that averaging and normalizing the staircases in the extended length is equivalent to plotting a cumulative distribution of dwell times (CDF) to the unfolding events, as shown in Fig.~\ref{fig:five}. Although the variability in the step lengths in the case of FLNa 16-23 may violate this equivalence, data analysis shows that they are the same. This result means that averaging over all the staircases gives a distribution of step lengths that is homogeneous over time. In addition, it is interesting to note that averaging over all the different responses gives rise to a smooth function. Having access to the single molecule data reveals that even complex processes at the molecular level average out in the bulk to give, in this case, two characteristic unraveling timescales, derived below.

Fitting each distribution then gives the functional form for the kinetics of unfolding~\citep{dwell}. The CDF of dwell times $F(t)$ and the normalized end-to-end length are constructed from staircases lasting at least  $2.75$ seconds to optimize the size of the data set and the accuracy of the fit. Since both constructs exhibit eight protein domains that are not identical, as well as domain-domain interactions in FLNa 16-23, each of the unfolding events should be associated with a specific rate constant. Surprisingly, we find that the empirical distributions for both constructs are well fit by a biexponential function with just two rate constants,  \begin{equation}
F(t)=1-Ae^{-a_{1}t}-(1-A)e^{-a_{2}t}
\end{equation}
where $a_{1}$ and $a_{2}$ are the rate constants and $A$ is the weight of the populations associated with each rate. We find that both rate constants are faster in the FLNa 16-23 construct, while $A$ remains the same between the polyproteins, as indicated in the figure legend. It is interesting to note that the two rate constants show a clear separation in timescales since $a_{1}$ is ten times slower than $a_{2}$ for both constructs. While the two rates may be associated with the weak and strong domains unfolding in linear FLNa 1-8, the molecular origin of the two populations is more complex in FLNa 16-23, as suggested by the broad step size distribution. For example, it is possible that the population that is slow to unravel represents interdomain interactions, while the faster population corresponds to domain unfolding. This simplified view of the process agrees with the observation that unfolding forces identified in FLNa 16-23 are weaker than those in FLNa 1-8. Even though the rate constants $a_1, a_2$ are on the same order of magnitude in both constructs, their differences lead to a more compliant response to force by FLNa 16-23, which releases $\sim70\%$ of its length after 2.75 seconds, while the more resilient FLNa 1-8 only releases $\sim45\%$.  

\section*{Conclusion}
Our results from force-extension and force-clamp data are indicative of domain-domain interactions between Ig domains in FLNa 16-23 that are not present in the linear FLNa~1-8. These domain-domain interactions lead to compact geometries and alter the pulling direction of each domain and thus lead to a broad distribution of extended lengths and rupture forces. While it is possible for misfolded structures to exhibit considerable mechanical stability~\cite{Oberhauser1999}, most often they display very weak mechanical resilience with no structural signatures~\cite{Garcia-Manyes2009}. Given that the folded structure of FLNa 16-23 has been solved~\cite{domain_1, domain_2} and that integrin binds to filamin under force in bulk assays~\cite{actin_mechanics1}, it is likely that the diversity of mechanical responses is due to natively folded polyproteins. The native globular structures involving multiple domain-domain interactions are stronger than the force of individual domain unfolding. These interactions sequester individual domains from the applied force, such that linear domain unfolding occurs after their rupture. As a result, they induce multiple timescales in the unfolding trajectories under a constant stretching force.

In Fig.~\ref{fig:six}, we show the linear and the quaternary structure of FLNa 16-23, inspired by the recent findings in \citep{domain_1, domain_2}. The maximum extension one could achieve by extending the folded chain is $\sim30$ nm in Fig.~\ref{fig:six} A. On the other hand, the shortest extension from rupturing domain-domain interactions is on the order of the length of a single domain of $\sim4$~nm, depending on the geometry of the arrangement with respect to the pulling direction, as shown by the arrows in Fig.~\ref{fig:six} B. Strong domain-domain interactions allow the protein to withstand higher pulling forces and shield the individual domains from the applied force. The released lengths then vary according to the topology of the compact globule. The multiplicity in the possible interactions offers an explanation for the broad range of forces we observe, while the number of distinct configurations accounts for the diversity in the end-to-end lengths.

Despite the complexity of possible polyprotein arrangements, we find a strong correlation between the force response of those events that are released after the rupture of a high force and those of individual Ig domains. We thus conclude that strong domain-domain interactions, such as those identified in Ig domains in titin~\cite{julio_titin}, need to be broken in order to unfold the linear polyprotein chain. It is interesting to note that ankyrin domains form supramolecular structures that behave like Hookean springs and rupture at similarly high forces~\citep{Lee}. 

Biologically, this broad range of interactions in the Filamin A dimer allows it to accommodate both weak and strong mechanical stresses in the environment. Experimentally it is clear that the non-actin binding domains have a higher complexity than those that bind to actin, which leads to their larger compliance in response to a stretching force and a scale-free distribution of rupture forces. It is plausible that domains 16-23 are the first to rearrange, or even unfold, in order to keep the stiffer cross-linked actin intact. Indeed, it has been shown that the partial rearrangement of the compact globule of FLNa 16-23 triggers biochemical pathways that may serve as part of the mechanical feedback system of the cell~\cite{protection}. The advantage of studying entire segments of Filamin A is that they portray the macroscopic response inside the cellular machinery. However, this is at the expense of knowing the specifics of the underlying molecular interactions. In future work, the complexity in Filamin A may be dissected by cloning specific pairs of domains to enumerate the mechanisms through which the protein responds to force.   

\section*{Acknowledgements} 
We would like to thank Michael Sheetz for seeding this project at the National University of Singapore and Lea-Laetitia Pontani for a careful reading of the manuscript. J.~B. holds a Career Award at the Scientific Interface from the Burroughs Wellcome Fund and was supported in part by the New York University Materials Research Science and Engineering Center Program of the National Science Foundation under Award Number DMR-0820341 and the National Science Foundation Career Award 0955621.
\bibliography{flnabib}	 

\clearpage

\begin{table}
\begin{center}
\resizebox{\columnwidth}{!} {
\begin{tabular}{|c|c|c|c|}
\hline
\textbf{FLNa} & \textbf{Range of the Number} & \textbf{Range of the Estimated} & \textbf{Range of the Estimated}\\
\textbf{Segment} & \textbf{of Amino Acids} & \textbf{Distance $\Delta L_{FX}$ (nm)} & \textbf{Step Size $\Delta L_{FC}$ at 100 pN }\\ \hline
1-8 & 93 - 103 & 30.71 - 34.41 & 25.69 - 28.78 \\ \hline
16-23 & 82 - 99 & 26.64 - 32.93 & 22.28 - 27.54 \\ \hline
\end{tabular}
}
\caption{Length characteristics of domains in FLNa 1-8 and 16-23. We list the number of amino acids (UniProt database), estimated range of distances and polypeptide chain extension at 100 pN for both constructs.}
\label{tab:table}
\end{center}
\end{table}

\clearpage
\section*{Figure Legends}

\subsubsection*{Figure 1}
A schematic of the dimer of FLNa monomers bound to actin filaments. The actin binding domain (ABD) anchors the FLNa protein to the actin. Linear repeats 1-15 exhibit strong interactions with actin, while FLNa 16-23 bridges the orthogonal actin filaments. Dimerization occurs between domains Ig 24 of two adjacent FLNa proteins. A zoom of the Ig domains 19-21 shows the domain-domain interactions that lead to the propeller-like architecture of the FLNa 16-23 construct. 
\subsubsection*{Figure 2}
A typical force-extension trace at a constant velocity of $400$~nm/s in (A) and a force-clamp trace at a constant force of $100$~pN in (B) for the linear construct FLNa 1-8. The sawtooth patterns is fit with the Worm-Like Chain (WLC) model (red line) to give the contour length $\Delta L_{FX}$ given out between consecutive peaks.  The regular staircase in force-clamp is indicative of individual Ig domain unfolding, as seen in tandem proteins, such as I27~\cite{i27,titin}.
\subsubsection*{Figure 3}
Distinct mechanical behaviors of FLNa 16-23: The diversity in the lengths unraveled is shown in (A), fit by the WLC model. A trajectory of length over time at a constant force of 100~pN is shown in (B), similarly revealing the release of diverse end-to-end lengths. Other traces under the same conditions reveal strong rupture events occurring before weak ones, shown in (C). The inset shows the decrease in the frequency of the number of consecutive weak events following a strong rupture. The observation of small after large peaks in force-extension measurements manifests itself as a separation of timescales in constant force experiments, as seen in (D). In this case the slow events occur prior to the fast ones.
\subsubsection*{Figure 4}
(A) A histogram of rupture forces for FLNa 16-23 under a constant velocity of $400$~nm/s. The shaded histogram represents previous constant velocity experiments~\cite{Furuike} investigating the Ig domain rupture in FLNa. Native FLNa 16-23 gives a much broader distribution than that of the Ig domains alone, which could be due to domain-domain interactions or other higher order structures. The power law decay in the inset captures the breadth of the distribution. (B) A histogram of contour lengths released between consecutive peaks for FLNa 16-23 is compared with the shaded histogram of Ig domain unfolding from the previous study~\cite{Furuike}. The inset shows a probability map of the histograms from Fig.~\ref{fig:four}A and \ref{fig:four}B. The map shows a high probability to observe $\Delta L_{FX}$ $\sim20-30$~nm, in the range of Ig domain unfolding, but a significant chance to originate from another mechanism. (C) A histogram of contour lengths released after the rupture of strong events also reveals a broad distribution. However, the inset shows a probability map of rupture force and $\Delta L_{FX}$ that isolates a peak at $\Delta L_{FX}$ at $\sim25-30$~nm, which indicates Ig domain rupture. (D) A histogram of step sizes for FLNa 16-23 under a constant force of $100$~pN reveals a broad distribution with a peak at $20$~nm,  indicative of Ig domain rupture. The inset shows  a uniform step size distribution for FLNa 1-8, which is peaked near the size of the Ig domains in the chain.
\subsubsection*{Figure 5}
Empirical CDFs for the dwell times and the normalized end-to-end length of FLNa 1-8 and 16-23 at a pulling force of $100$~pN. Data is fit with a biexponential function in Eq. (2) (FLNa 1-8: $a_{1}=0.11\pm0.03s^{-1}$ $a_{2}=1.9\pm0.5s^{-1}$ $A=0.8\pm0.1$; FLNa 16-23: $a_{1}=0.30\pm0.08s^{-1}$ $a_{2}=3.1\pm0.7s^{-1}$ $A=0.7\pm0.1$). The normalized length and CDF are almost identical in both constructs, which indicates no correlation between the step size and time of the rupture event.
\subsubsection*{Figure 6}
Examples of hypothetical arrangements of Ig domains 16-23. (A) Arrangement of the domains in a linear polyprotein chain. (B) The known interactions in FLNa 16-23 are incorporated into a globular configuration.

\clearpage

\clearpage

\clearpage
\begin{figure}
\begin{center}
\includegraphics[width=1.0\textwidth]{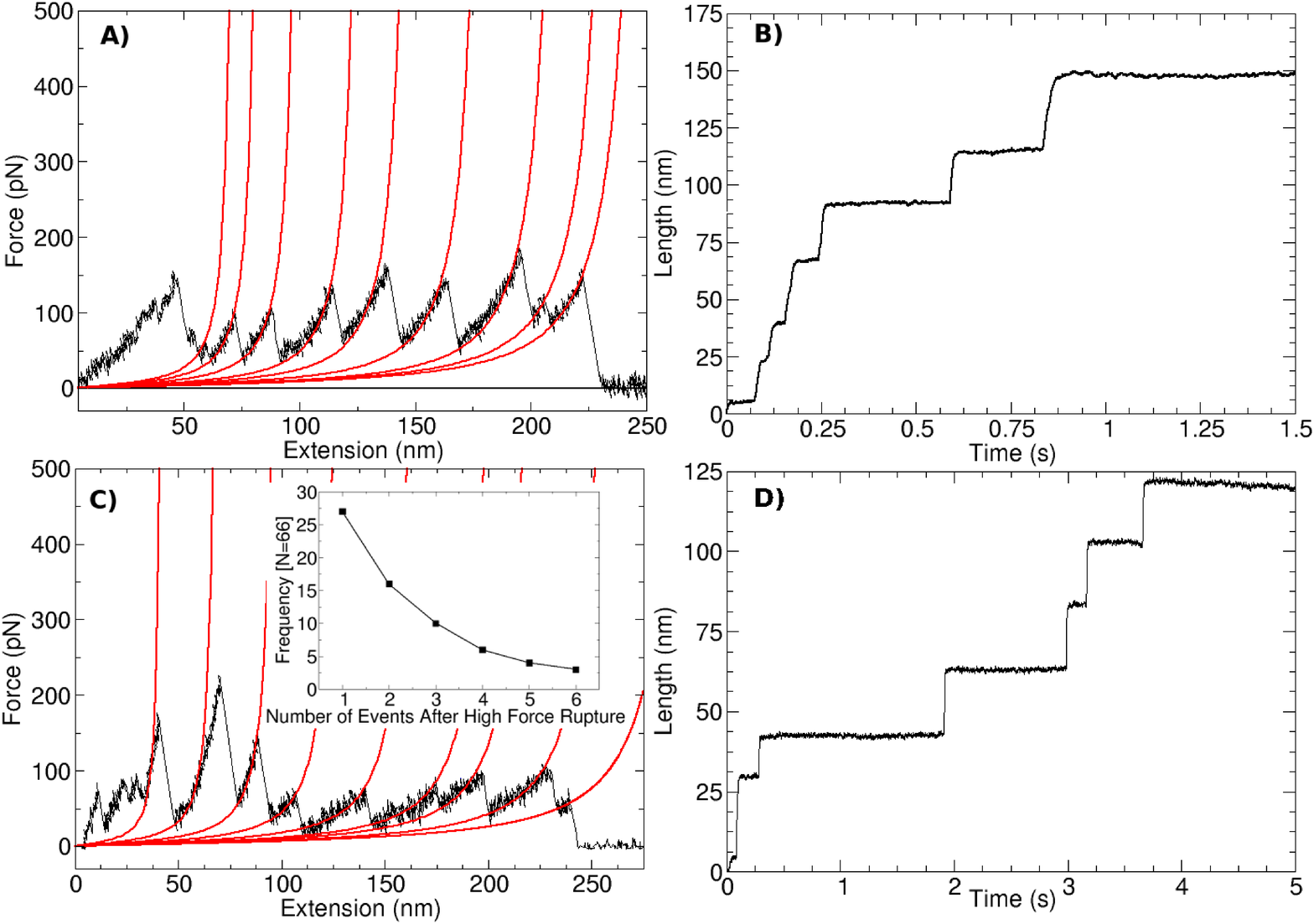}
\caption{}
\label{fig:three}
\end{center}
\end{figure}

\clearpage
\begin{figure}
\begin{center}
\includegraphics[width=1.0\textwidth]{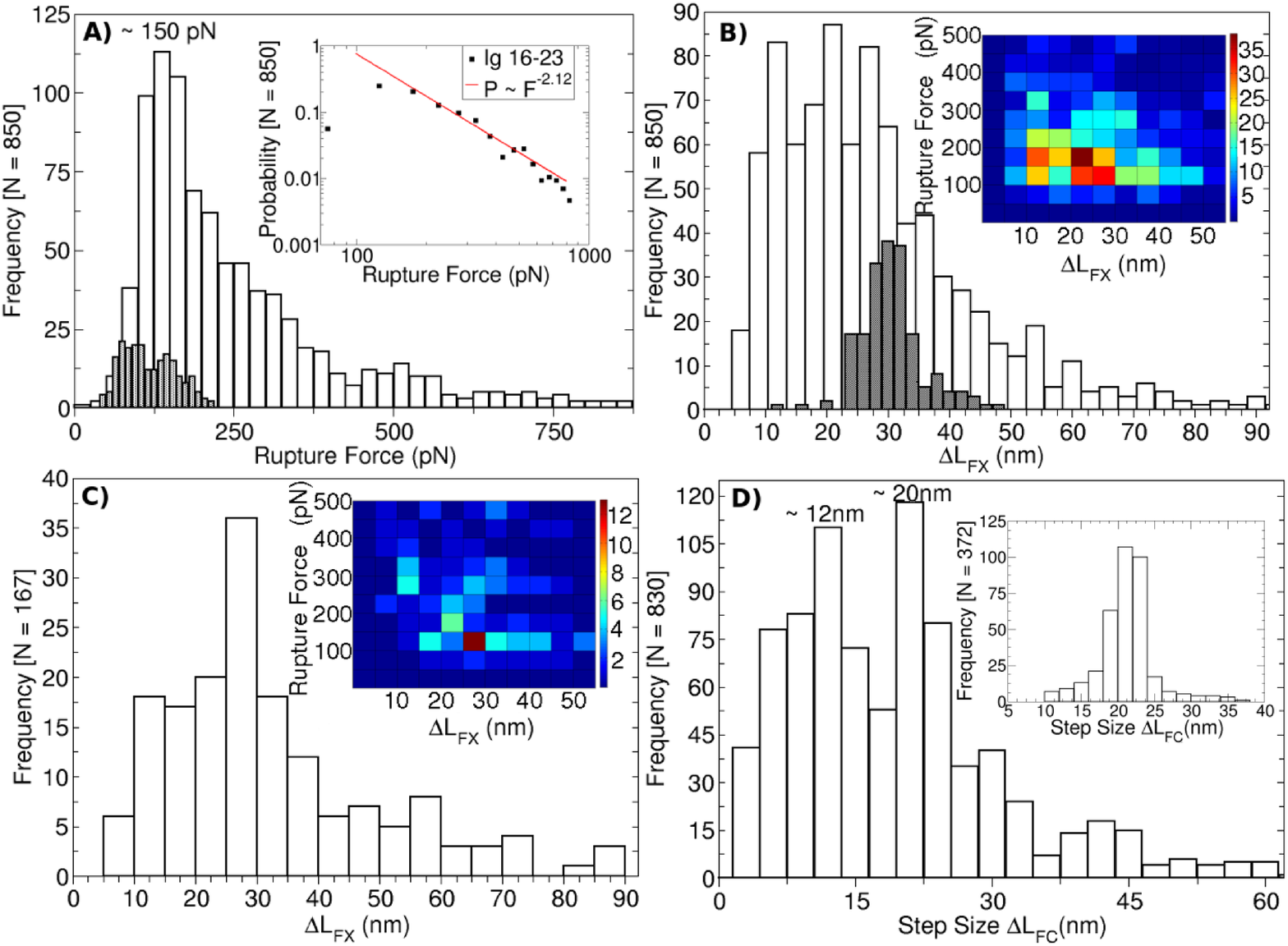}
\caption{}
\label{fig:four}
\end{center}
\end{figure}

\clearpage
\begin{figure}
\begin{center}
\includegraphics[width=1.0\textwidth]{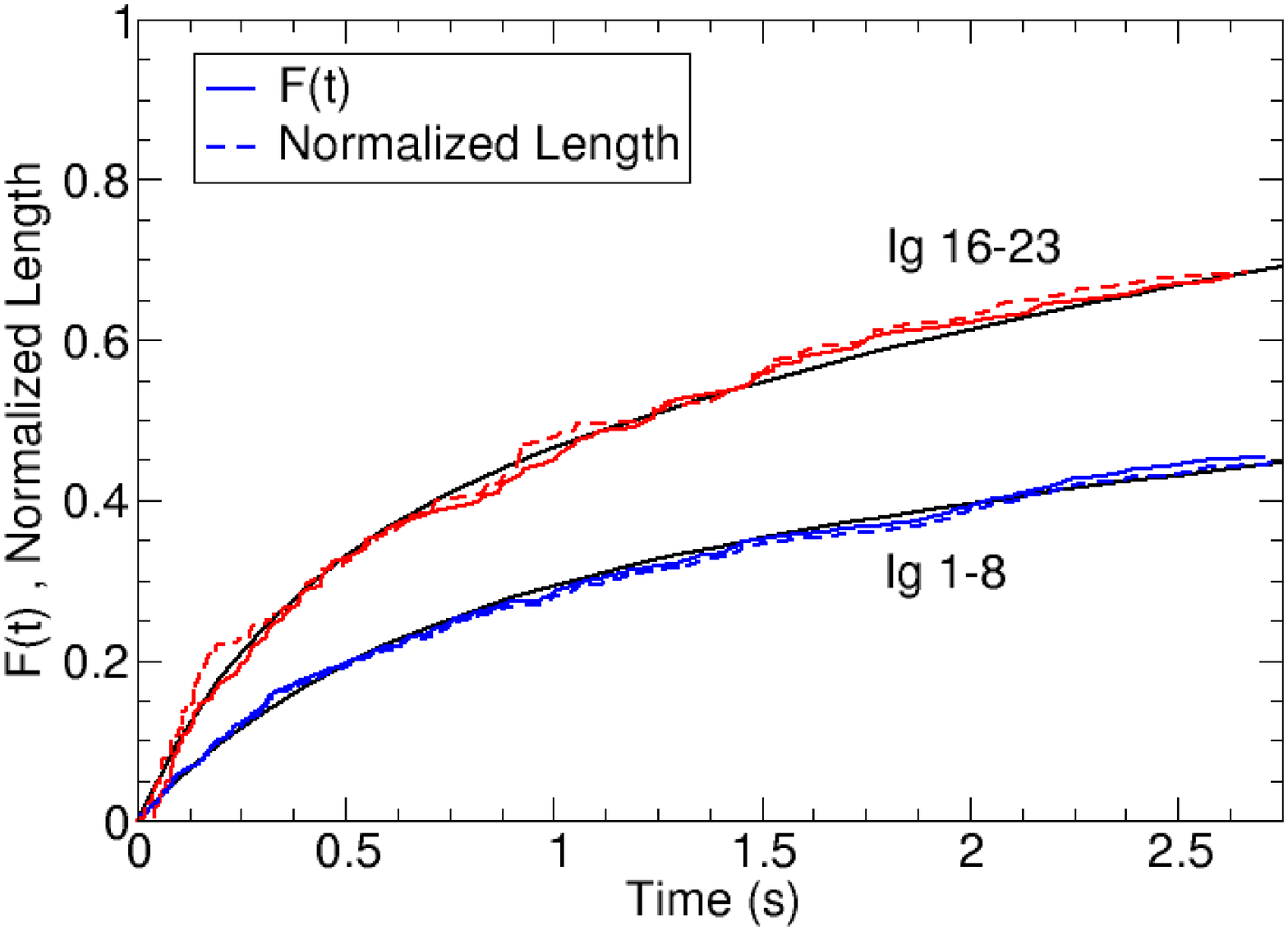}
\caption{}
\label{fig:five}
\end{center}
\end{figure}

\clearpage
\begin{figure}
\begin{center}
\includegraphics[width=1.0\textwidth]{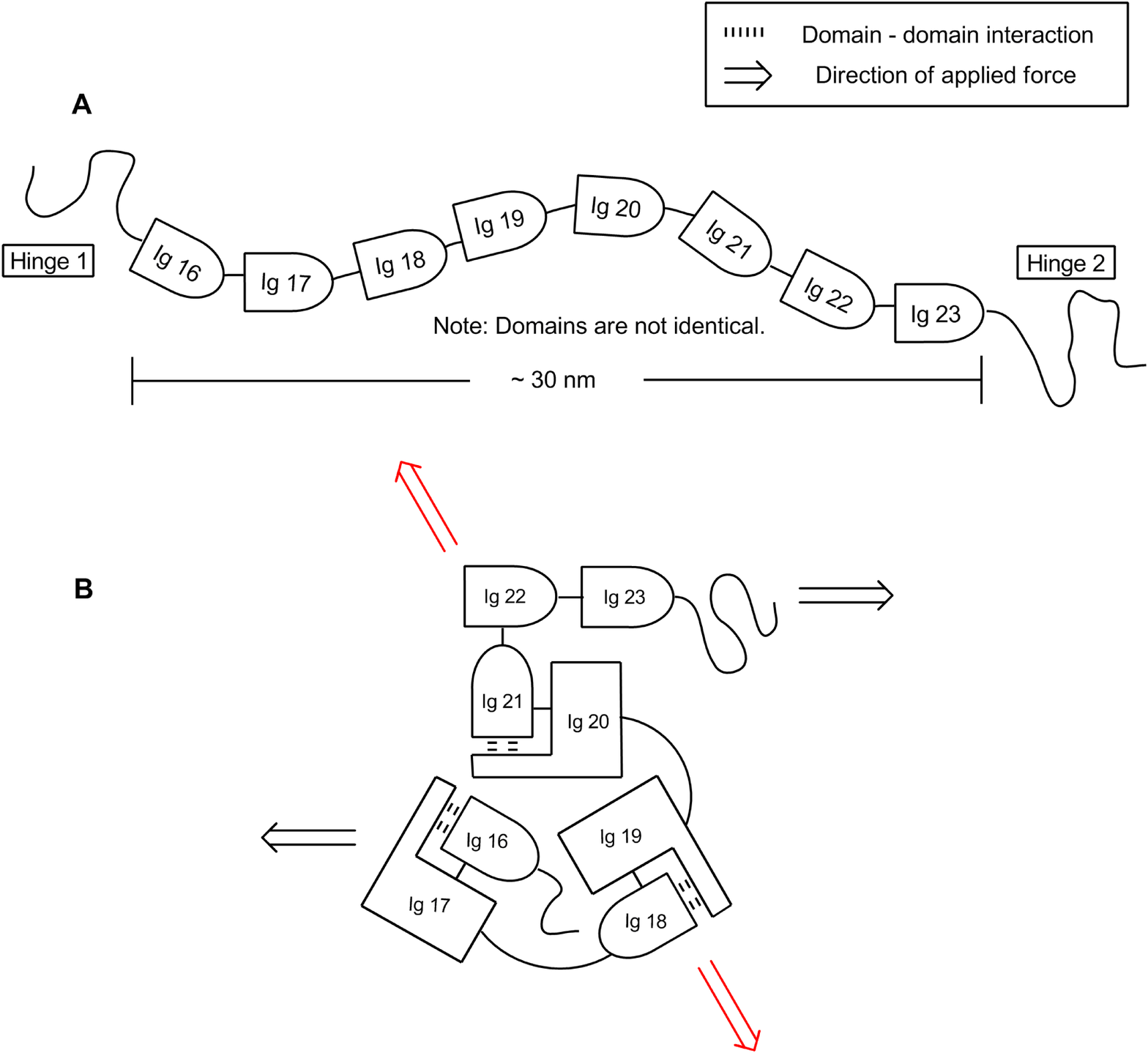}
\caption{}
\label{fig:six}
\end{center}
\end{figure}

\end{document}